\input amstex.tex
\input amsppt.sty   
\magnification 1200
\vsize = 9.5 true in
\hsize=6.2 true in
\NoRunningHeads        
\parskip=\medskipamount
        \lineskip=2pt\baselineskip=18pt\lineskiplimit=0pt
       
        \TagsOnRight
        \NoBlackBoxes

        \topmatter
        \title
        Integrals of products of Hermite functions     
        \endtitle
        \author
        W.-M.~Wang
        \endauthor
\address
Departement de Mathematique, Universite Paris Sud, 91405 Orsay Cedex, FRANCE;
\endaddress
        \email
{wei-min.wang\@math.u-psud.fr}
\endemail
{}
\abstract We compute the integrals of products of Hermite functions using the generating
functions. The precise asymptotics of products of $4$ Hermite functions are presented below. 
This estimate is relevant for the corresponding cubic nonlinear equation. 

\endabstract

        \bigskip\bigskip
        \bigskip
        
        \endtopmatter
        \bigskip
\document
\head{\bf 1. Introduction and statement of the theorem}\endhead
In this note, we compute the integrals of products of $4$ (normalized) Hermite 
functions:
$$W_{jpqk}=\int_{-\infty}^{\infty}h_j(x)h_p(x)h_q(x)h_k(x)dx,\quad j, p, q, k\geq 0\tag 1$$
using the generating functions.  The method applies to arbitrary products. The product of 
$4$ Hermite functions is motivated by the cubic nonlinearity in the equation:
$$i\frac\partial{\partial t}u
=\frac{1}{2}(-\frac{\partial^2}{\partial x^2}+x^2)u+|u|^2u.$$
The special case $p=q=0$ was computed in \cite{W}, where the author showed 
stability of the harmonic oscillator under the time dependent perturbation: 
$$V(x,t)=\delta|h_0(x)|^2\sum_{k=1}^\nu\cos(\omega_kt+\phi_k),\tag 2$$
for small $\delta$ and a set of frequencies $\omega=\{\omega_k\}$ close to full measure. 
In \cite {W}, the precise asymptotics:
$$W_{j00k}\sim\frac{1}{\sqrt{j+k}}e^{-\frac{(j-k)^2}{2(j+k)}}\quad \text {for } j+k\gg 1,\tag 3$$
played an essential role. Using (2)
$$\Vert Vh_k\Vert_2\leq\frac{1}{k^{1/4}}\to 0,\quad \text{ as } k\to\infty.$$
Hence the spatial part of the perturbation diminishes for higher Hermite modes contributing to stability.

 In this paper, we compute (1) for arbitrary $p$ and $q$. We prove 
 
 \proclaim{Theorem} 
 $$\aligned |W_{jpqk}|&\lesssim C_{p,q}\sum_{\ell=0}^{[\frac{p+q}{2}]}\frac{1}{\ell!}\big[\frac{(j-k)^2}{2(j+k)}\big]^{\ell}\cdot
 \frac{1}{ \sqrt{j+k}}e^{-\frac{(j-k)^2}{2(j+k)}}\\
&\lesssim \frac{C_{p,q}}{\sqrt{j+k}}e^{-\frac{(j-k)^2}{3(j+k)}}\qquad\quad\quad\text{ for }\frac{j-k}{\sqrt {j+k}}\geq\sqrt{p+q},\\
&\lesssim \frac{C_{p,q}}{\sqrt{j+k}}\qquad\qquad\qquad \qquad \text{ for all } j, k,\\
W_{jpqk}&=0\quad \qquad\qquad\qquad\qquad\qquad j+p+q+k \text{ odd},
  \endaligned\tag 4$$
where $[\,]$ denotes the integer part and $C_{p,q}\leq a^{p+q}$ for some $a>1$.
\endproclaim

We remark that except for the factor $C_{p,q}$, the estimate in (4) is essentially the same as in (3). 
In particular, the polynomial factor in front of the Gaussian is optimal and for fixed $p$ and $q$,
the Schur norm is of the same order as the operator norm.
Using (4), the result in \cite{W} extends immediately to potentials with exponentially decaying Hermite
coefficients. The rest of the paper is devoted to the proof of the Theorem. We first recall some basic
facts about the Hermite functions and the proof of (3).

The Hermite functions $h_j$ are the eigenfunctions of the harmonic oscillator:
$$\aligned Hh_j{\overset\text{def }\to =}&(-\frac{d^2}{d x^2}+x^2)h_j\\
=&\lambda_jh_j,\endaligned$$
with eigenvalues
$$\lambda_j=2j+1, \quad j=0,\,1..,$$
and 
$$h_j(x)=\frac{H_j(x)}{\sqrt {2^j j!}}e^{-x^2/2},\quad j=0,\,1...\tag 5$$
where $H_j(x)$ is the $j^{\text {th}}$ Hermite polynomial, relative to the weight
$e^{-x^2}$ ($H_0(x)=1$) and
$$\aligned &\int_{-\infty}^{\infty}e^{-x^2}H_j(x)H_k(x)dx\\
=&2^j j!\sqrt\pi\delta_{jk}.\endaligned\tag 6$$
So $$W_{jpqk}=\frac{1}{\sqrt{2^{j+p+q+k}j!p!q!k!}}\int_{-\infty}^{\infty}H_j(x)H_p(x)H_q(x)H_k(x)dx,\quad j, p, q, k\geq 0\tag 7$$

As in \cite{W}, the idea is to view the above integral as an $L^2$ product with the new
measure $e^{-2x^2}$ and reexpress the products of Hermite polynomials in $x$ as new Hermite polynomials
in $\sqrt 2 x$:
$$H_j(x)H_k(x)=\sum_{r=0}^{j+k}a_r H_r(\sqrt 2 x).\tag 8 $$
and similarly
$$H_p(x)H_q(x)=\sum_{\ell=0}^{p+q}b_\ell H_\ell(\sqrt 2 x).\tag 9$$
Using (8) and (9) in the integral in (7), and assuming (without loss of generality), $p+q\leq j+k$, we then have
$$I=\int_{-\infty}^{\infty}H_j(x)H_p(x)H_q(x)H_k(x)dx=\sum_{\ell=0}^{p+q}a_\ell b_\ell c_\ell,\tag 10$$
where $$c_\ell=\int_{-\infty}^{\infty}[H_\ell(\sqrt 2 x)]^2e^{-2x^2}dx=2^{\ell-\frac{1}{2}}\ell!\sqrt\pi.\tag 11$$

To find the coefficients $a_r$ and $b_\ell$, we use the generating functions as follows. 
Since 
$$\align&e^{2tx-t^2}=\sum_{n=0}^\infty\frac{t^n}{n!}H_n(x),\tag 12\\
&e^{2sx-s^2}=\sum_{m=0}^\infty\frac{s^m}{m!}H_m(x),\tag 13\endalign$$
which can be found in any mathematics handbook (cf. \cite {T} for connections with the
Mehler formula), multiplying (12, 13), we obtain 
$$\aligned &e^{2(t+s)x-(t^2+s^2)}=\sum_{n,m}\frac{t^ns^m}{n!m!}H_n(x)H_m(x)\\
=&e^{2(\frac{t+s}{\sqrt 2})\sqrt 2 x-(\frac{t+s}{\sqrt 2})^2}\cdot e^{-\frac{1}{2}(t-s)^2}\\
=&\sum_{\ell=0}^\infty H_\ell(\sqrt 2 x)\cdot\frac{(\frac{t+s}{\sqrt 2})^\ell}{\ell!}\cdot\sum_{r=0}^{\infty}
(-1)^r\frac{(t-s)^{2r}}{2^rr!}.\endaligned\tag 14$$
Using (14) in (8), we note that
$$\aligned a_0&=\frac{(-1)^{\frac{j-k}{2}}}{2^{\frac{j+k}{2}}(\frac{j+k}{2})!}\cdot (j+k)!,\qquad\quad j+k\text{ even}\\
&=0\qquad\qquad\qquad\qquad\qquad\qquad\text{otherwise},\endaligned\tag 15$$
by taking $2r=j+k$, which is the only contributing term. In \cite{W}, we computed the case $p=q=0$:
$W_{j00k}=a_0b_0c_0$, which we recall below.

\proclaim{Lemma}
$$\aligned W_{jk}{\overset\text{def }\to =}W_{j00k}=&\int_{-\infty}^{\infty}h_0^2(x)h_j(x)h_k(x)dx\\
=&\frac{(-1)^{\frac{j-k}{2}}}{2^{j+k}\sqrt{j!k!}}\cdot\frac{(j+k)!}{(\frac{j+k}{2})!}\sqrt{\frac{\pi}{2}}\qquad j+k \text{ even},\\
=&0\qquad\qquad\qquad\qquad\qquad\qquad\quad\text{otherwise}.\endaligned\tag 16$$
Let $$J=\frac{j+k}{2},\, K=\frac{j-k}{2},\tag 17$$
assuming $j\geq k$, without loss. When $J\gg 1$,
$$\align W_{jk}&=\left[1+\Cal O\left(\frac{1}{J}\right)\right]\frac{(-1)^{\frac{j-k}{2}}J!}{\sqrt{2 J(J+K)!(J-K)!}},\tag 18\\
|W_{jk}|&\leq\frac{1}{\sqrt J}e^{-K^2/{2J}}.\tag 19\endalign$$
\endproclaim
\demo {Proof} (16) follows directly from (15, 7). We only need to obtain the asymptotics in (18, 19). This is an
exercise in Stirling's formula:
$$j!=\big(\frac{j}{e}\big)^j\sqrt{2\pi j}(1+\frac{1}{12j}+\frac{1}{288j^2}+...)$$
or its log version
$$\log j!=(j+\frac{1}{2})\log j-j+\log\sqrt{2\pi}+...\tag 20$$
Here it is more convenient to use the latter. Using (17, 20),
$$\aligned &\log\frac{(j+k)!}{2^{j+k}(\frac{j+k}{2})!}=\log\frac{(2J)!}{J!}-\log 2^{2J}\\
= &J\log J-J+\frac{1}{2}\log J+\Cal O(J^{-1}).\endaligned\tag 21$$
So $$\frac{(j+k)!}{2^{j+k}(\frac{j+k}{2})!}=\left[1+\Cal O\left(\frac{1}{J}\right)\right]\frac{J!}{\sqrt {\pi J}},$$ 
using (21). Hence 
$$W_{jk}=\left[1+\Cal O\left(\frac{1}{J}\right)\right]\frac{(-1)^{\frac{j-k}{2}}J!}{\sqrt{2 J(J+K)!(J-K)!}},\quad J\gg 1$$
which is (18). Using the fact that
$$j!=\sqrt{2\pi j}\left(\frac{j}{e}\right)^j e^{\lambda_j}$$
with $$\frac{1}{12j+1}<\lambda_j<\frac{1}{12j},\text { for all }j\geq 1,$$
and applying the inequalities (with $x=K/J$):
$$\phi(x){\overset\text{def }\to =}(1+x)\log (1+x)+(1-x)\log (1-x)\geq x^2$$
for all $x\in[0,1)$ and $\phi(x)\geq ax^2$ with $a>1$ for $x\in[7/10, 1)$,
we obtain (19). (When $x=K/J=1$, (19) follows by a direct computation using Stirling's formula.)
\hfill $\square$
\enddemo

\head{\bf 2. Proof of the theorem}\endhead 

From (14), the $t^js^k$ term in the RHS is among the terms:
$$\aligned &\sum_{\ell=0,\,\ell\sim j+k}^{j+k} H_\ell(\sqrt 2 x)\cdot\frac{(t+s)^\ell}{2^{\frac{\ell}{2}}\ell!}
\big[(-1)^r\frac{(t-s)^{2r}}{2^rr!}\big ]\big |_{2r+\ell=j+k}\\
=&\sum_{\ell=0,\,\ell\sim j+k}^{j+k} H_\ell(\sqrt 2 x)\cdot\frac{(t+s)^\ell}{2^{\frac{\ell}{2}}\ell!}
(-1)^{\frac{j+k-\ell}{2}}\frac{(t-s)^{j+k-\ell}}{2^{\frac{j+k-\ell}{2}}({\frac{j+k-\ell}{2}})!},
\endaligned\tag 22$$
where $\ell\sim j+k$ means that $\ell$ has the same parity as $j+k$. Equating the coefficients in
front of the $t^js^k$ term in the second line of (14) and (22), we obtain 
$$\aligned H_j(x)H_k(x)&=j!k!\sum_{\ell=0,\,\ell\sim j+k}^{j+k}H_\ell(\sqrt 2 x)\cdot
\frac{(-1)^{\frac{j+k-\ell}{2}}}  {2^{\frac{j+k}{2}} \ell! (\frac{j+k-\ell}{2})! }\\
&\qquad\qquad\qquad\qquad\qquad\sum_{r=0}^\ell(-1)^{k-r}C_\ell^rC_{j+k-\ell}^{k-r}\\
&=\sum_\ell a_\ell H_\ell(\sqrt 2 x)
\endaligned\tag 23$$
from (8). Similarly
$$\aligned H_p(x)H_q(x)&=p!q!\sum_{i=0,\,i\sim p+q}^{p+q}H_i(\sqrt 2 x)\cdot
\frac{ (-1)^{\frac{p+q-i}{2} }} {2^{\frac{p+q}{2}} i! (\frac{p+q-i}{2})!}\\
&\qquad\qquad\qquad\qquad\qquad\sum_{m=0}^i(-1)^{q-m}C_i^mC_{p+q-i}^{q-m}\\
&=\sum_ib_i H_i(\sqrt 2 x)
\endaligned\tag 24$$
from (9).

So 
$$\aligned a_\ell&=\frac{(-1)^{\frac{j+k-\ell}{2}}j!k!} {2^{\frac{j+k}{2}} \ell! (\frac{j+k-\ell}{2})! }
\sum_{r=0}^\ell(-1)^{k-r}C_\ell^rC_{j+k-\ell}^{k-r},\\
&\qquad\qquad\text{ for } \ell=j+k, j+k-2,... (\ell\geq 0),\\
&=0\qquad\qquad\text{ otherwise},\endaligned\tag 25$$
and 
$$\aligned b_\ell&=\frac{(-1)^{\frac{p+q-\ell}{2}}p!q!} {2^{\frac{p+q}{2}} \ell! (\frac{p+q-\ell}{2})! }
\sum_{m=0}^\ell(-1)^{q-m}C_\ell^mC_{p+q-\ell}^{q-m},\\
&\qquad\qquad\text{ for } \ell=p+q, p+q-2,... (\ell\geq 0),\\
&=0\qquad\qquad\text{ otherwise}.\endaligned\tag 26$$

Using (25, 26) in (10, 7), assuming $p+q+j+k$ even, otherwise $I=0$, we then have
$$\aligned W_{jpqk}
=\frac{\sqrt{j!k!p!q!}\sqrt{\frac{\pi}{2}}}{2^{j+p+q+k}}&\sum_{\ell=0,\,\ell\sim p+q}^{p+q}
\frac{(-1)^{\frac{j+k+p+q}{2}-\ell} 2^\ell}{(\frac{j+k-\ell}{2})! (\frac{p+q-\ell}{2})!\ell!}
(\sum_{r=0}^\ell(-1)^{k-r}C_\ell^rC_{j+k-\ell}^{k-r})\\
&\qquad\qquad\qquad(\sum_{m=0}^\ell(-1)^{q-m}C_\ell^mC_{p+q-\ell}^{q-m}),\qquad j+p+q+k\text { even}.\endaligned
\tag 27$$
For each $\ell=p+q, p+q-2, ..., (\ell\geq 0)$, we then need to estimate
$$I_{jk}^{(\ell)}=\frac{\sqrt{j!k!}}{2^{j+k}}
\frac{1}{(\frac{j+k-\ell}{2})!}\sum_{r=0}^\ell(-1)^{k-r}C_\ell^rC_{j+k-\ell}^{k-r}.\tag 28$$
and
$$I_{pq}^{(\ell)}=\frac{\sqrt{p!q!}}{2^{p+q}}
\frac{1}{(\frac{p+q-\ell}{2})!}\sum_{m=0}^\ell(-1)^{q-m}C_\ell^mC_{p+q-\ell}^{q-m}.\tag 29$$
Then
$$W_{jpqk}=\sqrt{\frac{\pi}{2}}\sum_{\ell=0,\,\ell\sim p+q}^{p+q}
(-1)^{\frac{j+k+p+q}{2}-\ell}\frac {2^\ell}{\ell!}I_{jk}^{(\ell)}I_{pq}^{(\ell)},\qquad p+q\leq j+k.\tag 30$$
We check that when $p+q=0$, the sum in (30) reduces to the term $\ell=0$ and
$$W_{j00k}=\frac{(-1)^\frac{j+k}{2}}{2^{j+k}\sqrt{j!k!}}\cdot\frac{(j+k)!}{(\frac{j+k}{2})!}\sqrt{\frac{\pi}{2}},\qquad j+k \text{ even},$$
same as in (16).
\smallskip
\noindent{\it Estimates on} $I_{jk}^{(\ell)}$.

We first look at (28). When $\ell>0$, we need to perform the sum over $r$. Rewrite
$$\aligned I_{jk}^{(\ell)}&=\frac{\sqrt{j!k!}}{2^{j+k}}
\frac{1}{(\frac{j+k-\ell}{2})!}\sum_{r=0}^\ell(-1)^{k-r}\frac{\ell!}{r!(\ell-r)!}\frac{(j+k-\ell)!}{(k-r)!(j-\ell+r)!}\\
&=\frac{(-1)^k}{2^{j+k}}\cdot \frac{(j+k-\ell)!}{(\frac{j+k-\ell}{2})!}\cdot\frac{\ell!}{\sqrt{j!k!}}
\sum_{r=0}^\ell(-1)^{r}\frac{k!}{r!(k-r)!}\frac{j!}{(\ell-r)!(j-\ell+r)!}\\
&=(-1)^kF_1F_2, \qquad \ell=p+q, p+q-2, ... (\ell\geq 0),\endaligned\tag 31$$
where $F_2$ denotes the sum. 

We note that 
$$I_{jk}^{(0)}=\frac{1}{2^{j+k}\sqrt{j!k!}}\cdot\frac{(j+k)!}{(\frac{j+k}{2})!}\sim \frac{1}{\sqrt {j+k}}e^{-\frac{(j-k)^2}{2(j+k)}}$$
from the Lemma. So we write 
$$ \aligned F_1&= I_{jk}^{(0)}\cdot \frac {\ell!}{2^\frac{\ell}{2}}\cdot \frac {(j+k-\ell-1)!!} {(j+k-1)!!}\\
|F_1|&\lesssim |I_{jk}^{(0)}|\cdot \frac {\ell!}{2^\frac{\ell}{2}}\cdot\frac{e^{\frac{\ell}{2}}}{(j+k)^{[\frac{\ell}{2}]}} .\endaligned
\tag 32$$
$F_2$ can be written as 
$$\aligned F_2&=\sum_{r=0}^\ell(-1)^{r}\frac{k!}{r!(k-r)!}\frac{j!}{(\ell-r)!(j-\ell+r)!}\\
&=\sum_{r=0}^\ell (-1)^{r}C_k^rC_j^{\ell-r}\\
&=\sum_{r=0}^{[\ell/2]}(-1)^{r}C_k^rC_{j-k}^{\ell-2r},\qquad j\geq k, j+k\geq \ell,\endaligned\tag 33$$
since $(t-s)^k(t+s)^j=(t^2-s^2)^k(t+s)^{j-k}$ and both expressions in (33) give the coefficients
in front of the $s^\ell t^{j+k-\ell}$ term ($\ell\leq j+k$). So 
$$|F_2|\leq \sum_{r=0}^{[\ell/2]}\frac{k^r}{r!}\frac{(j-k)^{\ell-2r}}{(\ell-2r)!}.$$

We also need to estimate $I_{pq}^{(\ell)}$, $\ell=p+q, p+q-2,... (\ell>0).$ Since
$p+q\leq j+k$, we use norm estimates. Comparing (29) with (26), we have 
$$I_{pq}^{(\ell)}=\frac{(-1)^{\frac{p+q-\ell}{2}}\ell! }{2^{\frac{p+q}{2}}\sqrt{p!q!}}b_\ell,\tag 34$$
where $b_\ell$ is as defined in (24). From (24)
$$\aligned b_\ell\int &H_\ell^2(\sqrt 2 x)e^{-2x^2} dx=\int H_p(x)H_q(x)H_\ell(\sqrt 2 x)e^{-2x^2} dx\\
&\leq\big[\int H_p^2(x)H_q^2(x)e^{-2x^2}dx\big]^{1/2}\big[\int H_\ell^2(\sqrt 2 x)e^{-2x^2} dx\big]^{1/2}\endaligned$$
So $$|b_\ell|\leq 2^{\frac{p+q}{2}}\frac{\sqrt{p!q!}}{2^{\frac{\ell}{2}}\sqrt{\ell!}},\tag 35$$
where we used the normalization conditions in (5, 6) and the $L^\infty$ estimate \cite{T}
$$\Vert h_p\Vert_{L^\infty}\leq \frac{1}{p^{1/12}}<C.$$
Using (35) in (34), we then have 
$$|I_{pq}^{(\ell)}|\leq\sqrt{\frac{\ell!}{2^\ell}}.$$

So the terms in the sum in (30) can be estimated as follows:
$$\aligned&\frac {2^\ell}{\ell!}|I_{jk}^{(\ell)}I_{pq}^{(\ell)}|\\ 
\leq &|I_{jk}^{(0)}|e^{\frac{\ell}{2}}\sqrt{\ell!}\sum_{r=0}^{[\ell/2]}\big(\frac{j-k}{\sqrt{j+k}}\big)^{\ell-2r}\cdot\big(\frac{k}{j+k}\big)^r
\cdot\frac{1}{r!(\ell-2r)!}\\
\leq &|I_{jk}^{(0)}|e^{\frac{\ell}{2}}\sqrt{\ell!}\sum_{r=0}^{[\ell/2]}\frac{1}{2^r}\cdot X^{\ell-2r}\cdot\frac{1}{r!(\ell-2r)!}\\
\leq &\frac{1}{\sqrt{j+k}}e^{-\frac{X^2}{2}}e^{\frac{\ell}{2}}X^\ell \sum_{r=0}^{[\ell/2]}\frac{\sqrt{\ell!}}{(2X^2)^r}\cdot\frac{1}{r!(\ell-2r)!},\endaligned\tag 36$$
where $X=\frac{j-k}{\sqrt{j+k}}$. Using Stirling's formula to relate $\sqrt{\ell!}$ and $(\ell-2r)!$ to
$\frac{\ell}{2}!$ and $(\frac{\ell}{2}-r)!$, we have 
$$(36)\leq \frac{C^\ell}{\sqrt {j+k}}\sum_{r=0}^{[\ell/2]}e^{-\frac{X^2}{2}}\cdot\frac{(\frac{X^2}{2})^{\frac{\ell}{2}-r}}{(\frac{\ell}{2}-r)!}\cdot\frac{(\frac{\ell}{2})!}{r!(\frac{\ell}{2}-r)!},$$
for some $C>1$.  Summing over $r$ and then $\ell$, we obtain the Theorem. 
\hfill $\square$

\enddocument
\end